\documentclass[11pt,a4paper,fleqn]{article} \textheight=25 true cm \textwidth=17.5 true cm
\usepackage{epsfig}
\usepackage{graphicx}
\usepackage{amsmath}
\usepackage{amssymb}
\usepackage{cite}
\usepackage{cleveref}
\usepackage{rotating}
\usepackage{color}
\usepackage[colorlinks]{hyperref}
\usepackage{setspace}
\begin{document}
\begin{center}
{\Large{\textbf{Singly excited \textit{S-}states of compressed two-electron ions}}}\\[8 mm]
J. K. Saha$^{1}$, S. Bhattacharyya$^{2}$, T. K. Mukherjee$^{3,*}$\\[4 mm]
$^{1}$Indian Association for the Cultivation of Science, Jadavpur, Kolkata 700032, India\\
$^{2}$Acharya Prafulla Chandra College, New Barrackpore, Kolkata 700131, India\\
$^{3}$Narula Institute of Technology, Agarpara, Kolkata 700109, India \\
*E-mail : drtapanmukherjee@gmail.com\\
\end{center}
\begin{abstract}
A detailed analysis on the effect of spherical impenetrable confinement on the structural properties of two-electron ions  in $ S- $states have been done. The energy values of $1sns$ [$n=2-4$] ($^3S^e$) states of helium-like ions ($Z = 2-5$) are estimated within the framework of Ritz variational method by using explicitly correlated Hylleraas-type basis sets. The correlated wave functions used here are consistent with the finite boundary conditions due to spherical confinement. A comparative study between the singlet and triplet states originating from a particular electronic configuration shows incidental degeneracy and the subsequent level-crossing phenomena. The thermodynamic pressure felt by the ion inside the sphere pushes the energy levels towards continuum. Critical pressures for the transition to strong confinement regime (where the singly excited two-electron energy levels cross the corresponding one-electron threshold) as well as for the complete destabilization are also estimated.
\end{abstract}
\vspace{-0.5cm}
\section{Introduction}
Studies on the broad subject area of confined atomic systems have been started soon after the advent of quantum mechanics. A paradigm shift in the facility for doing controlled experiment on confined atom and modern high-speed computational resource opens up a new horizon for this subject. The modification of the structural and spectral properties of the atom or the molecule depending upon the number and degree of confining parameters is the interesting aspect of the problem both theoretically and experimentally. Comprehensive reviews \cite{sab09,can08,sen14} are now available on this topic. Theoretically, different types of confinement model can be realized depending on different physical situations \textit{e.g.} atoms under plasma environment \cite{sil09,bhat15}, endohedrally confined atoms and molecules in fullerene cages \cite{xu96}, impurities in quantum dots or nano crystals \cite{gen10,nor08}, matter under extreme pressure in zeolite sieves \cite{jac97} or in the walls of nuclear reactors \cite{wal00} \textit{etc}. Moreover the confined atom models assume contemporary significance in understanding the cores of Jovian planets such as Jupiter and Saturn \cite{gul99}. Vast applicability of the outcomes from such studies in different branches of science and technology makes the subject a topic of immense interest in the recent times. \\
Hydrogen atom confined in impenetrable spherical cavity was first studied by Michels \textit{et al.} \cite{mic37} to model the effects of pressure on it’s energy and polarizability. Subsequently, many workers \cite{som38,mont0211,lau02,lau04,cif09,ste08,aqu9511} have addressed this problem from different angles to develop a comprehensive understanding about such confined systems. Helium atom confined by such spherical cavity is a much complicated numerical problem as compared to the hydrogen-like ions for obvious reasons. For ions having two or more electrons, the effect of electron correlation plays an important role in the structure of the wavefunctions and energy levels. This effect also undergoes some modifications inside the confinement and it is extremely necessary to adopt an appropriate methodology so that the correlation is properly taken care of. In order to achieve high numerical accuracy in structure calculations of helium-like systems in non-relativistic regime, it is well known that the trial wave function should be expanded in Hylleraas type basis set (an explicit function of inter-electronic distance $r_{12}$) consistent with the Dirichlet's boundary condition and the basis integrals are to be evaluated within a finite domain to incorporate the effect of impenetrable confinement. Recently Bhattacharyya \textit{et al.} \cite{bhat15,bhat13} solve the integral analytically in such a way that the problem of linear dependency for larger dimension is clearly avoided even for very strong confinement regime. As a result, the correlated Hylleraas basis set becomes flexible for a complete range of confining parameters and produces accurate energy levels for helium-like ions. Although there exists several investigations \cite{bhat13, gim67, lud79, mar91, jos92, aqu03, aru06, flo08, lau09, flo10, mont13} on the series of $ ^1S^e $ states of helium-like systems, the progress of studies on the series of $ ^3S^e $ states of the same are rather scanty \cite{aru06, flo08,flo10, mont13,pat04,yak11,mont15}. To be specific, Patil and Varshni \cite{pat04} first estimated the energy values of $ 1s2s$ ($^3S^e $) state of helium under the perturbative framework with a simple single-term uncorrelated wavefunction satisfying proper boundary condition for $R \longrightarrow \infty$ and the electron-electron interaction was replaced by an effective screening of the nuclear charge. Banerjee \textit{et al.} \cite{aru06} also performed the calculation for $ 1s2s$ ($^3S^e $) state of helium by adopting more accurate variational technique with two-parameter correlated wavefunction. Flores-Riveros and Rodríguez-Contreras \cite{flo08} first adopted 25-term correlated generalized Hylleraas basis (with $e^{-\gamma r_{12}}$ term) to estimate the energy values of lowest triplet $ 1s2s $ state of helium-like ions ($ Z=2-5 $). Later, Flores-Riveros \textit{et al.} \cite{flo10} used 20 term generalized Hylleraas function in the framework of Ritz variational technique as well as with perturbative method for $ 1s2s$ ($^3S^e $) state of helium to show that the Hylleraas function under variational method produces much accurate results for the energy levels compared to the perturbative method. It is to mention that due to the presence of $ e^{-\gamma r_{12}} $ factor in the generalized Hylleraas basis, the solution of the generalized eigenvalue equation within a finite domain becomes extremely difficult for larger dimensions due to linear dependency problems. Yakkar \textit{et al.} \cite{yak11} carried out the estimations of energy levels of box confined helium-like atoms ($ Z=1-4 $) using a combination of quantum genetic algorithm (QGA) and Hartree-Fock Roothaan (HFR) method. They have used uncorrelated two-electron wavefunction and thus the estimated energy values are less accurate as compared to other results \cite{flo08,flo10}. Montogomery and Pupyshev \cite{mont13} expanded the wavefunction in the permetric Hylleraas coordinates to estimate the energy values of $ 1sns$ ($^3S^e $) states ($ n=2-5 $) of helium and in a very recent work \cite{mont15} they estimated the energy levels of $ 1s2s $ and $ 1s3s $ states of helium and lithium for few large box radius values. It is thus to be noted that none of these works are aimed at precise determination of the energy values for the strong confinement region \textit{i.e.} small values of the box radius. As a result the accurate value of the ionization radius is missing. Thus a systematic analysis of energy levels for $ ^3S^e $ states of helium-like ions with accurate extended explicitly correlated Hylleraas basis function is a need of time.\\
In the present work, we have performed a systematic computation to determine the accurate non-relativistic energy levels of $1sns$ [$n=2-4$] ($^3S^e$) states of helium-like atoms [$Z=2-5$] inside an impenetrable spherical cavity of radius $R$. Accuracy of the estimated energy eigenvalues have been tested by studying the convergence pattern of the eigen energies with respect to a systematic increment in the number of terms ($N$) in the basis sets. The effect of spatial confinement on these energy levels have been studied by varying the confinement radius ($R$). The energy values gradually become positive as $ R $ decreases. Due to spatial restriction imposed upon the wave function, the ion inside the cavity will feel a pressure which increases with decrease of $ R $. The critical thermodynamic pressure ($ P_c $) have been estimated at the destabilization limit of the two-electron ion inside the impenetrable cavity of critical radius ($ R_c $). We have made further analysis of the results in combination with the energy values of $1sns$ [$n=1-4$] ($^1S^e$) states of the same ions \cite{bhat13} under consideration. `\textit{Incidental degeneracy}' \cite{bhat15,sen05,dut15} and subsequent `\textit{level-crossing}' phenomenon between the excited singlet and triplet $ S- $states originating from the same electronic configuration have been observed. In order to estimate the variation of ionization potential (IP) and the ionization potential depression (IPD) as a function of $R$, we have also calculated the energy eigenvalues of $ns$ ($^2S$) [$n = 1-2$] and $2p$ ($^2P$) states of the H-like sub-systems of the two-electron ions. We note that as $R$ decreases, both the two-electron excited states as well as the respective one-electron threshold move towards destabilization which result in reduction of IP. It is remarkable that below a certain value of $R$, the two-electron energy levels move above the respective one-electron threshold and become \textit{quasibound}. We have made an effort to calculate the amount of pressure at which the $1sns$($^1S^e$)[$n=1-4$] states and $1sn's$ [$n=2-4$] ($^3S^e$) states of two-electron ions [$Z=2-5$] when the system goes to high confinement region and become quasibound. The remaining part of the article is organized in the following manner: in Section 2, we describe the present theoretical technique; Section 3 is devoted to the discussion of the results and finally, the conclusion is given in Section 4.
\section{Method}
The non-relativistic Hamiltonian (in a.u.) of a spatially confined two-electron ion can be written as
\begin{eqnarray}\label{1}
H=\sum^{2}_{i=1}\left[-\frac{1}{2}\nabla_{i}^{2}-\frac{Z}{r_i}\right]+\frac{1}{r_{12}}+V_R(r_1,r_2)
\end{eqnarray}
Here $Z$ is the nuclear charge and we assume that the atom/ion is placed at the center of an impenetrable spherical cavity of radius $R$ where the nature of the potential $V_R(r_1,r_2)$ is expressed as
\begin{eqnarray}\label{2}
V_R (r_1,r_2)&=&0 ~~~~~~~~~~ 0 \leq (r_1,r_2) \leq R    \nonumber \\
             &=& \infty ~~~~~~~~ otherwise
\end{eqnarray}
The structure of the potential as given by equation (2) imposes a spatial restriction on the system, and therefore, can be used to model the confinement due to pressure. The orbital wave function $\Psi$ satisfies the boundary condition 
\begin{eqnarray}\label{eq:ab}
\Psi(r)=0~~~~~~~~~~~~~at~~~ r \geq R
\end{eqnarray}
The Schrodinger equation $H\Psi=E\Psi$ is to be solved to obtain the energy eigenvalues where the wave function is subject to the normalization condition $\langle \Psi \vert \Psi \rangle=1$ within the sphere.\\
Exploiting the translational and rotational invariance of the two-electron Hamiltonian (eqn. 1), the variational equation for $^{1,3}S^{e}$ states originating from two $s$-electrons ($1sns$ configuration) in Hylleraas co-ordinates ($r_1, r_2, r_{12}$; $r_1, r_2$ being the distances of the two-electrons from the nucleus and $ r_{12}=|r_1-r_2| $) is given by \cite{tkm94}, 
\begin{eqnarray*}\label{eq:ab1}
\delta\int \left[\frac {1}{2}\left(\frac{d\Psi_S}{dr_1}\right)^2+ \frac {1}{2}\left(\frac{d\Psi_S}{dr_2}\right)^2+\left(\frac{r_1^2-r_2^2+r_{12}^2}{2r_1r_{12}}\right)\left(\frac{d\Psi_S}{dr_1}\right)\left(\frac{d\Psi_S}{dr_{12}}\right)\right.\nonumber~~~
\end{eqnarray*}
\begin{equation}
\left.+\left(\frac{r_2^2-r_1^2+r_{12}^2}{2r_2r_{12}}\right)\left(\frac{d\Psi_S}{dr_2}\right)\left(\frac{d\Psi_S}{dr_{12}}\right)+(V-E)\Psi_S^2\right]dV_{r_1,r_2,r_{12}}~=~0
\end{equation}
For a `\textit{free}' system, where no external environment affects the atom/ion, the upper limit of integration for $r_1$ and $r_2$ is infinity whereas, in the present case, the upper limit is $R$. The upper and lower limits of integration for $r_{12}$ are $(r_1 + r_2)$ and $|r_1 - r_2|$, respectively. The volume element is expressed as 
\begin{eqnarray}
dV_{r_1,r_2,r_{12}}~=~r_1 r_2 r_{12} dr_1 dr_2 dr_{12}
\end{eqnarray}
The correlated function must be consistent with the boundary condition given by eq. (3) and can be written as 
\begin{eqnarray}\label{eq:ab2}
\Psi_S(r_1,r_2,r_{12})=(R-r_1)(R-r_2)F(r_1,r_2,r_{12})
\end{eqnarray}
with
\begin{eqnarray}\label{eq:ab2}
F(r_1,r_2,r_{12})=(1\pm \hat{P}_{12})e^{-\sigma_1 r_1-\sigma_2 r_2}\sum_{l\geq 0}\sum_{m\geq 0}\sum_{n\geq 0}C_{lmn}r_1^l r_2^m r_{12}^n
\end{eqnarray}
$\hat{P}_{12}$ is the permutation operator for two electrons. The upper sign in the \textit{r.h.s.} of eq. (7) is taken for singlet states whereas the lower sign is for triplet states. $\sigma_1$ and $\sigma_2$ are the nonlinear parameters taking care of the effect of radial correlation in the wave function whereas the angular correlation effect is incorporated through different powers of $r_{12}$. The total number of parameters (\textit{N}) in the basis set is defined as the total number of different $(l, m, n)$ sets (eq. 7) taken in the expansion of $F(r_1,r_2,r_{12})$. We have used the Nelder-Mead algorithm \cite{nel65} to optimize the nonlinear parameters \textit{i.e.} $\sigma$'s in eq. (7). The linear variational parameters \textit{i.e.} $C_{lmn}$'s along with the energy eigenvalues $E$ are obtained by solving the generalized eigenvalue equation
\begin{eqnarray}
\underline{\underline{H}}~\underline{C}=E~\underline{\underline{S}}~\underline{C}
\end{eqnarray}
where $\underline{\underline{H}}$ is the Hamiltonian matrix, $\underline{\underline{S}}$ is the overlap matrix and $\underline{C}$ is the column matrix consisting of linear variational parameters. The wave function is  normalized for each confining radius \textit{R} to account for the reorientation of charge distribution inside the sphere. All computations are carried out in quadruple precision to ensure numerical accuracy for extended Hylleraas basis sets within a finite domain.\\
The variational equation for $nl$ ($^2L$) states of one electron ion within a cavity of radius $R$ can be written as
\begin{eqnarray}\label{11}
\delta\int_{0}^{R} \left[\frac{1}{2}\left\{\left(\frac{\partial f}{\partial r}\right)^{2}+\frac{l(l+1)}{r^2}f^2\right\}+\left\lbrace V_{R}(r)-E\right\rbrace f^{2}\right]r^{2}dr=0
\end{eqnarray}
where, the one-particle effective potential $V_R(r)$ is taken from eq. (2). The radial function $f(r)$ is given by
\begin{eqnarray}
f(r)=(R-r)r^k\sum_{i}C_{i}e^{-\rho_{i}r}
\end{eqnarray}
where $k=0$ and 1 for $^2S$ and $^2P$ states respectively. In this calculation, we have taken 21 different nonlinear parameters ($\rho_{i}$'s) in a geometrical sequence $\rho_{i}=\rho_{i-1}\gamma$, $\gamma$ being the geometrical ratio \cite{bhat15}. Such choice of non-linear parameters enables us to cover the full region of space in a flexible manner by adjusting $\gamma$. The energy values $E$'s and linear co-efficients $C_{i}$'s are determined from eq. (8).\\
The truncation of wavefunction at a finite distance (eq. 3) imposes a thermodynamic pressure upon the ions inside the sphere which increases with decrease of $R$. We have calculated the pressure felt by the helium-like ions inside the sphere by using the first law of thermodynamics. Under an adiabatic approximation, the pressure ($ P $) inside the impenetrable cavity can be expressed as \cite{bhat15,bhat13}
\begin{eqnarray}
P=-\frac{1}{4 \pi R^2}\frac{dE_R}{dR}\simeq -\frac{1}{4 \pi R^2}\frac{\bigtriangleup E_R}{\bigtriangleup R}
\end{eqnarray}
where $ E_R $ is the ground state energy of the ion inside the sphere of radius $ R $. In the present calculation, we have taken $ \bigtriangleup R =1.0 \times 10^{-5}$ and calculated the corresponding $ \bigtriangleup E_R$ around a particular $ R $ to calculate the pressure inside the cavity. 
\section{Results and discussions}
The expansion length $N=l+m+n$ (in eq. 7) of the Hylleraas basis set is varied in a systematic fashion to study the convergence pattern of the energy eigenvalues of $1sns$ ($^3S^e$) [$n=2-4$] states of helium-like ions ($Z = 2-5$) for each value of the confinement radius $R$ up to the limit of destabilization. The results for such convergence behavior are demonstrated in table 1 where the energy values of $1s2s$ ($^3S^e$) state of helium corresponding to different confining radii ($R$) are given. Table 1 shows that the expansion length of the wavefunction up to $N = 161$ satisfying $l + m + n=10$ is sufficient to converge the energy eigenvalues at least up to the eighth decimal place. We have obtained a similar convergence pattern for all the other ions also. However, for higher excited states, we observe slow convergence and, therefore,  in subsequent tables, we have given the energy values up to sixth decimal place only. It is important to mention here that the present $1s2s$ ($^3S^e$) state energy value of the compressed helium atom converges up to the eighth decimal place for the confinement radius near the destabilization limit. The above observation ensures that the present method can deal with extended correlated basis sets to yield sufficiently accurate energy values within the complete range of finite spherical confinement. With this expansion length, we have estimated the energy values of singly excited $1s2s$ and $1s3s$ ($^3S^e$) states for the ions having $ Z=2-5 $.\\
The electron correlation contributes directly to the energy values of two-electron ions and, therefore, is extremely important for the accuracy of such structure calculations. The present methodology enables us to quantify the contribution of angular correlation on the energies of two-electron ions. In order to asses the contribution of angular correlation to the energy of the lowest lying triplet $1s2s$ state of helium, we have estimated the energy eigenvalues ($E_{unco}$) with no $r_{12}$ terms in the basis \textit{i.e.} by setting $ n=0 $ in eq. (7) to form an uncorrelated basis. Subsequently, we have estimated the percentage contribution of angular correlation to the energy as $f=\frac{\Delta E}{E}\times 100\%$ (where, $\Delta E$ = $E-E_{unco}$) for different values of $R$ and plotted in figure-1. It is evident from figure-1 that $f$ increases in a rapid rate w.r.t. $R$ near the destabilization region. To be specific, $f=0.04\%$ for $R=100$ a.u. and finally it reaches a value of 3.62\% for $R=2.18$ a.u. \textit{i.e.} close to the destabilization limit of the $1s2s$($ ^3S^e $) state of helium. We have noted that this feature is consistent for other two-electron ions also and thus clearly reveals the importance of considering angular correlation in the basis.\\
Table 2 furnishes a consolidated view of the $1sns$ ($^3S^e$) [$n=2-4$] state energies of He within a spherical box of different radii whose range is appropriately chosen to yield the energies of the `\textit{free}' system as well as those approaching the destabilization limit. We observe that the energy of the confined helium atom in $1s2s$ ($^3S^e$) state is almost unaltered and equal to the `\textit{free}' system when the radius is 20.0 a.u. or more. The effect of the confinement starts to become prominent when the radius is further reduced. For higher excited states like $1s3s$ and $1s4s$ ($^3S^e$) states, the confinement effect becomes prominent at a higher value of $ R $. The results have been compared with currently available data \cite{aru06,flo10,mont13,pat04,yak11}. In general, for each excited state at each radius of confinement, the present energy values are either the lowest yet obtained or exactly the same as the available results. The energies of spherically confined Li$^+$, Be$^{2+}$ and B$^{3+}$ ions in $1sns$ ($^3S^e$) ($n = 2-4$) states as displayed in tables $ 3-5 $ respectively are being reported for the first time in literature.\\
It is evident from tables $ 2-5 $ that as the radius ($R$) of the impenetrable cavity decreases, the ions become less bound and thus the number of excited states gradually decreases. For example, the $1s4s$ ($^3S^e$) state destabilizes below 5.1 a.u. whereas the $1s2s$ and $1s3s$ ($^3S^e$) state survive till $ R= 3.6$ and 2.18 a.u. respectively. The variation of energy eigenvalues of $1sns$ ($^3S^e$) [$n=2-4$] states of \textit{He} \textit{w.r.t.} the confining radii ($R$) is depicted in figure-2(a). It is evident from figure-2(a) that the energy values remain almost unaltered for a range of $R$ and below that, rapidly approaches towards the destabilization limit producing a `\textit{knee}' around some particular value of \textit{R} which increases for higher excited states. This `\textit{knee}' clearly shows the region of $ R $ from where the effect of confinement becomes crucial. Similar feature is observed for all other ions also. A zoomed view of figure-2(a) near the strong confinement (\textit{i.e.} small $R$) region is given in figure-2(b). We have plotted the energies of $1s2s$ ($^3S^e$) states of all the ions ($Z = 2-5$) in figure-3(a) and a zoomed view of the destabilization region in figure-3(b) and observe general consistency. The ion with greater nuclear charge survives to a lower value of the confining radius ($R$). It is also evident that for the respective H-like ions, the $2s$ state is fragmented much before $1s$ with the decrease of $R$.\\
Ionization potential (IP) for $1sns~(^3S^e)$ state of a two-electron ion may be defined as the amount of energy required to ionize the outer electron from the $1sns~(^3S^e)$ state. We can calculate the IP for $1sns~(^3S^e)$ state by taking the difference of the two-electron energy level and the corresponding one-electron energy level ($ 1s $). With decrease of $ R $, all the energy levels are modified and hence, IP changes \textit{w.r.t.} $ R $. It is observed from tables $ 2-5 $ for all the ions that with decrease of $R$, IP decreases and below certain value of $R$, the two-electron energy levels move above the one-electron threshold. Ionization potential depression (IPD) may be estimated from the difference of IP's within and without (\textit{i.e.} free case) the impenetrable confinement. In figure-4, we have plotted the IP and IPD for He in $1s2s~(^3S^e)$ state as a function of $R$.\\
One of the most interesting features of the atomic systems confined under different potentials is the evolution of quasi-bound states \cite{car09}. These states are structurally similar with the discrete two-electron bound states but are found embedded in the one-electron continuum. Such states have also been observed experimentally \cite{cap92}. For a two-electron ion, the ground state and all singly excited energy levels, in general, lie below the first ionization threshold. Tables 2-5 show that for high values `$R$' (\textit{i.e.} almost `\textit{free}' system), this feature is  maintained for all the ions. As $R$ decreases, all singly excited states of two-electron ions become less bound more rapidly than the respective one-electron ion and as a result, the two-electron energy levels move above the one-electron threshold at some particular value of $ R $ to become quasi-bound. For a better understanding, the energy level diagrams of helium for $R=40.0$ a.u. and $R=10$ a.u. are given in figure-5(a) and 5(b) respectively. We observe that, for $R=40.0$ a.u., the singly excited $1sns$ [$n=2-4$] ($^3S^e$) energy levels of helium lie below the He$^{+}$ ($1s$) level. For $R = 10.0$ a.u., $1s3s$ and $1s4s~(^3S^e)$ states of helium move above the $1s$ threshold and lie below 2\textit{s} threshold. Thus both the $1s3s$ and $1s4s~(^3S^e)$ levels can now (\textit{at $ R=10.0 a.u. $}) be categorized as quasi-bound states. In this quasi-bound situation, it seems as if the inner electron quantum number of the singly excited two-electron state has been changed from 1 to 2. At $R<3.0$ a.u. we observe well-converged energy level of $1s2s~(^3S^e)$ state of helium lying above He$^{+}$ ($1s$) threshold and no upper one-electron continuum exists as both the levels $2s$ and $2p$ of He$^{+} $ are destabilized. Similar feature is obtained for other ions also.\\
For a `\textit{free}' two-electron ion, the $1s2s~(^3S^e)$ state is energetically more negative than the $1s2s~(^1S^e)$ state. For example, the $1s2s~(^3S^e)$ level of He lies below $1s2s~(^1S^e)$ level \cite{bhat13} for the whole range of $R$ and the corresponding variations are given in figure-6(a). We have plotted a similar comparative diagram in figure-6(b) for singlet and triplet \textit{S}-states origination from $1s3s$ configuration of He. It is interesting to note that, although $1s3s$-triplet level of He lies below the singlet level for large confinement radius, the gap between them gradually decreases as $ R $ decreases. It is seen in figure-6(b) that, below some value of $R$ ($ < 4.0 $ a.u.), a `\textit{level crossing}' phenomenon takes place \textit{i.e.} $1s3s~(^3S^e)$ state moves above the $1s3s~(^1S^e)$ level. Ultimately the triplet $ 1s3s $ state of helium destabilizes at $R=3.6$ a.u. whereas the singlet $ 1s3s $ state survives down to $ R=3.208 $ a.u. These results show that an `$incidental~ degeneracy$' has taken place for $1s3s~(^3S^e)$ and $1s3s~(^1S^e)$ states of He around $R \sim 4.0$ a.u. and then the `$level~ crossing$' occurs between two states of same electronic configuration and total angular momentum but having different spin multiplicity. Similarly, for $1s4s~(^3S^e)$ and $1s4s~(^1S^e)$ states of He, $incidental~ degeneracy$ and subsequent $level~ crossing$ are observed at a value of $R$ lying somewhere between 5.5 and 6.5 a.u. The phenomenon of $incidental~degeneracy$ was reported in case of shell-confined hydrogen atom by Sen \cite{sen05}. More recently, Dutta \textit{et al.} \cite{dut15} and Bhattacharyya \textit{et al.} \cite{bhat15} have noted such phenomenon in case of slowly moving hydrogen-like ion in quantum plasma and in case of two-electron ions embedded in strongly coupled plasma respectively where two initially non-degenerate states are brought to a same energy level by adjusting external parameters.\\
In Coulomb potential, the hydrogen-like ions are expected to show degenerate energy levels for all \textit{l-values} corresponding to a particular principal quantum number \textit{n}. The present work observes the removal of this \textit{l-}degeneracy when the whole system is subjected to a confinement although the potential inside the sphere is purely Coulombic. We can see from table-2 that, for $ He^+ $ ion, the $2s$ and $2p$ energy levels are at the same level of $ -0.5 $ a.u. for high values of confining radius $ R$, \textit{i.e.} from from $ R=100.0 $ a.u. to $ 20.0 $ a.u. Evidently, the one-electron ion $ He^+ $ behaves almost like a `\textit{free}' ion within this range of $ R $. When $ R $ decreases to $ 15.0 $ a.u. or below, the energy values of $2s$ and $2p$ levels become more positive, but in a different rate and hence, the \textit{l-}degeneracy is lifted. The $ 2s $ levels becomes positive more rapidly than the $ 2p $ level as $ R $ decreases. The role of confinement in the removal of \textit{l-}degeneracy in a similar fashion is observed for other hydrogen-like ions also and the results are given in tables $ 3-5 $.\\
As a result of the truncation of the wave function at a finite distance due to the confinement, the ions inside the impenetrable sphere experience a thermodynamic pressure. This pressure is actually responsible for modification of different structural properties of two-electron ions such as crossing of one-electron ionization threshold and evolution of quasi bound states, destabilization of the three-body system at high confinement \textit{etc.} We have made an effort to calculate the value of confinement radius ($ R_{th} $) and corresponding pressure ($ P_{th} $) when the energy level of a two-electron ion crosses the first ionization threshold as well as the `\textit{critical}' radius ($ R_c $) and pressure ($ P_c $) near the destabilization region experienced by the ions under consideration for both singlet and triplet states upto $ 1s4s $ configuration by using eq. (11). The results are displayed in tables 6 and 7 respectively. It is to mention that the thermodynamic pressure can be calculated by using eq. (11) only when the system is in the ground state. For higher excited states, this relation does not hold as the equilibrium criteria is not satisfied because of finite lifetimes of such states. However, we can assume that the amount of pressure would be same for all the states at a particular value of the confinement radius. Therefore, the pressure \textit{felt} by an ion in an excited state for a particular value of $ R $ can be estimated by calculating the energy gradient of the ground state $ 1s^2~(^1S^e) $ \textit{w.r.t.} $ R $ around the same vale of $ R $ and then by applying eq. (11). For example, the $ 1s2s~(^1S^e) $ state of helium moves above the $ N=1 $ ionization threshold of $ He^+ $ when the confinement radius ($ R_{th} $) is 5.33 a.u. We calculate the energy gradient of $ 1s^2~(^1S^e) $ state of helium \textit{w.r.t.} $ R $ around $ R=5.33 $ a.u. and get the pressure ($ P_{th} $) for $ 1s2s~(^1S^e) $ state of helium from eq. (11). In a similar fashion, all other pressures for different states crossing the first ionization threshold and for destabilization of different three-body ions have been calculated. However, for $ 1s4s~(^{1,3}S^e) $ states, the crossing of the ionization threshold occurs at a very high radius. As a consequence, the ground state energy at this high radius is almost equal to the `\textit{unrestricted}' ion and therefore, the energy gradient is almost zero and calculation of pressure is extremely difficult. This study reveals that highly excited states of two-electron ions becomes quasi-bound even at a very feeble confinement. \\
For the sake of completeness of the study, we have also studied the behavior of $ 1s2s $($ ^{1,3}S^e $) states of $ H^- $ ion ($ Z=1 $). It is well known that there is only one bound state (\textit{i.e.} $ 1s^2 $) of a `\textit{free}' $ H^- $ ion below the first ionization threshold. It is remarkable that both the singlet and triplet states of $ H^- $ ion are found to exist in presence of pressure confinement which is consistent with Ref.\cite{mont15}. The variation of energy eigenvalues of $ 1s2s $($ ^{1,3}S^e $) states of $ H^- $ ion within pressure confinement are depicted in figure-7. It is seen that the triplet state lies energetically below the singlet state. Both the states move towards destabilization as `R' decreases and the triplet state survives up to a smaller value of `R' than the singlet state. This feature is similar to the $ 1s2s $($ ^{1,3}S^e $) states of other ions considered here as can be seen from figure-6(a).
\section{Conclusion}
The energy levels of atoms under high pressure generated due to spatial confinement can be studied effectively using the variational method with Hylleraas basis sets yielding very consistent and highly accurate atomic data under confinement. The methodology is extendable to study atomic
structure under confinement with different internal potentials. Application of this methodology to study the transition properties of atoms under spatial confinements is likely to be an interesting problem for future investigations.\\\\
\textbf{{Acknowledgments}}\\
SB acknowledges the financial support under grant number PSW-160/14-15(ERO) from University Grants Commission, Govt. of India. TKM acknowledges the financial support from the Department of Atomic Energy, BRNS, Govt. of India under grant number 37(3)/14/27/2014-BRNS.

\newpage
\begin{table}[tbp]
\caption {\rm {Convergence of energy values ($-E$ a.u.) of the $1s2s$ ($^3S^e$) state of helium with respect to number of terms (N) in wave function within the impenetrable spherical cavity of radius $R$ a.u.}}
\begin{center}
\begin{tabular}{l c c c c c c}\\
\hline\hline\vspace{-0.2cm}\\
 & \multicolumn{6}{c}{$-E$ (in a.u.)}\\
\cline{2-7}\vspace{-0.2cm}\\
 $N$    & $R=5.0$ & $ R=3.0$ & $R=2.5$ & $R=2.4$ & $R=2.308$ & $R=2.18$\\
\hline\vspace{-0.2cm}\\
3   & 2.037 780 56 & 1.351 900 29 & 0.712 491 93 & 0.529 139 33& 0.362 294 80&0.011 566 45 \\
7  &  2.045 930 72 & 1.370 027 44& 0.749 395 89& 0.562 538 72& 0.366 793 68& 0.036 968 57\\
13  & 2.047 364 37&   1.370 356 37& 0.751 504 78& 0.565 246 29& 0.366 972 15& 0.038 192 56\\
22  & 2.048 036 88&   1.370 486 82& 0.751 634 77& 0.565 407 04& 0.367 032 22& 0.038 221 16\\
34  & 2.048 042 98 & 1.370 508 44& 0.751 655 16& 0.565 422 35& 0.367 049 50& 0.038 239 12\\
50  & 2.048 043 99&  1.370 510 42& 0.751 656 63& 0.565 423 97& 0.367 050 26 & 0.038 239 27\\
70  & 2.048 044 07&  1.370 510 60& 0.751 656 67& 0.565 424 02& 0.367 050 32& 0.038 240 26\\
95  & 2.048 044 09&  1.370 510 61& 0.751 656 67& 0.565 424 02& 0.367 050 32&0.038 240 28 \\
125&  2.048 044 09&  1.370 510 61& 0.751 656 67& 0.565 424 02& 0.367 050 32& 0.038 240 29\\
161&  2.048 044 09 & 1.370 510 61& 0.751 656 67& 0.565 424 02& 0.367 050 32& 0.038 240 29\\
\hline\hline
\end{tabular}
\end{center}
\begin{center}
\end{center}
\end{table}

\begin{table}[tbp]
\caption {\rm {Energy values of $ He $ and $He^+$ confined in a spherical cavity of radius R. The uncertainty of the calculated energy values is of the order of $10^{-6}$ a.u.}}
\begin{center}
\begin{tabular}{l l l l l l l l}\\
\hline\hline\vspace{-0.2cm}\\
$R$ & \multicolumn{3}{c}{$-E$ (in a.u.) of $ He $}&&\multicolumn{3}{c}{$-E~$ (in a.u.) of $He^+$}\\
\cline{2-4}\cline{6-8}\vspace{-0.2cm}\\
(a.u.)& $1s2s$ & $1s3s$ & $1s4s$ && $1s$ &$ 2s $&$ 2p $\\
\hline\vspace{-0.2cm}\\
40.0 & 2.175 227 & 2.068 543 & 2.036 312 && 2.000 000 & 0.500 000 & 0.500 000 \\
20.0 & 2.175 227 & 2.066 814 & 2.005 650 && 2.000 000 & 0.500 000 & 0.500 000 \\
15.0 &2.175 195  & 2.055 436 & 1.940 623 && 2.000 000 & 0.499 999 & 0.500 000 \\
     &2.174 5$^{a}$    &           &           &           &           &     \\     
10.0 &2.172 627  & 1.979 435 & 1.692 000 && 2.000 000 & 0.499 948 & 0.499 978 \\
     &2.164 7$^{a}$    &1.979 424$^{c}$ &1.691 923$^{c}$&           &           &     \\
     &2.171 4$^{b}$    &           &           &           &           &      \\
     &2.172 627$^{c}$\\
     &2.171 46$^{d}$  &           &           &           &           &           \\
     &2.172 6$^{e}$\\
9.0  &2.169 481        &1.935 386  & 1.571 962 && 2.000 000 & 0.499 765 & 0.499 898 \\
     &2.157 0$^{a}$    & 1.935 383$^{c}$& 1.571 914$^{c}$ &           &           &           \\
     &2.168 3$^{b}$    &           &           &           &           &           \\
     &2.169 481$^{c}$\\
8.0  &2.162 784        &1.866 425  & 1.395 165 && 2.000 000 & 0.499 001 & 0.499 553 \\
     &2.142 9$^{a}$    & 1.866 425$^{c}$ & 1.395 143$^{c}$ &           &           &           \\
     &2.161 7$^{b}$    &           &           &           &           &           \\
     &2.162 784$^{c}$  &  & &           &           & \\
     &2.162 8$^{e}$\\
7.0  &2.148 564  & 1.754 413 & 1.123 624 && 2.000 000 & 0.496060  & 0.498 162 \\
     &2.116 6$^{a}$    & 1.754 413$^{c}$ & 1.123 615$^{c}$ &           &           & \\
     &2.147 7$^{b}$    &           &           &           &           &           \\
     &2.148 564$^{c}$  &           &           &           &           &           \\
     &2.147 48$^{d}$  &           &           &           &           &           \\     
6.0  &2.117 816        &1.562 909  & 0.683 444 && 2.000 000 & 0.485 663 & 0.49 3020 \\
     &2.065 8$^{a}$    & 1.562 909$^{c}$ & 0.683 439$^{c}$ &           &           &\\
     &2.117 1$^{b}$    &           &           &           &           &           \\
     &2.117 816$^{c}$  &           &           &           &           &           \\
     &2.117 8$^{e}$    &           &           &           &           &           \\
5.5  &2.090 241        &1.415 329  & 0.356 858 && 2.000 000 & 0.473 379 & 0.486 794 \\
5.4  &           &           & 0.279 473 && 2.000 000 & 0.469 924 & 0.485 030 \\
5.3  &           &           & 0.197 455 && 2.000 000 & 0.466 038 & 0.483 042 \\
5.19 &           &           & 0.102 096 && 2.000 000 & 0.461 201 & 0.480 563 \\
5.101&           &           & 0.027 875 && 1.999 998 & 0.456 804 & 0.478 305\\
5.1  &           &           &           && 1.999 998 & 0.456 752 & 0.478 279 \\
\hline\hline\\
\multicolumn{6}{l}{$^{a}$Ref.\cite{pat04}; $^{b}$Ref.\cite{aru06}; $^{c}$Ref.\cite{mont13}; $^{d}$Ref.\cite{yak11};  $^{e}$Ref.\cite{flo08}}&\\
\end{tabular}
\end{center}
\end{table}
\begin{table*}[tbp]
Table 2 continued
\begin{center}
\begin{tabular}{l l l l l l l l}\\
\hline\hline\vspace{-0.2cm}\\
$R$ & \multicolumn{3}{c}{$-E$ (in a.u.) of $ He $}&&\multicolumn{3}{c}{$-E~$ (in a.u.) of $He^+$}\\
\cline{2-4}\cline{6-8}\vspace{-0.2cm}\\
(a.u.)& $1s2s$ & $1s3s$ & $1s4s$ && $1s$ &$ 2s $&$ 2p $\\
\hline\vspace{-0.2cm}\\
5.0  &2.048 044  &1.211 382  &           && 1.999 997 & 0.451 225 & 0.475 438 \\
     &1.961 5$^{a}$    &1.211 382$^{c}$  &           &           &           &           \\
     &2.047 3$^{b}$    &           &           &           &           &           \\
     &2.048 044$^{c}$  &           &           &           &           &           \\
     &2.047 87$^{d}$  &           &           &           &           &           \\
     &2.048 0$^{e}$    &           &           &           &           &           \\
     &2.048 044$^{f}$\\
4.5  &1.981 851  &0.921 954  &           && 1.999 983 & 0.411 339 & 0.454 909 \\
4.0  &1.874 612  &0.497 750  &           && 1.999 900 & 0.338 955 & 0.417 800 \\
     &1.727 7$^{a}$    &0.497 750$^{c}$   &           &           &           &           \\
     &1.873 4$^{b}$    &           &           &           &           &           \\
     &1.874 612$^{c}$  &&           &           &           &       \\
     &1.873 31$^{d}$  &           &           &           &           &           \\
     &1.874 6$^{e}$    &           &           &           &           &           \\ 
3.9  &1.845 780        &0.389 989  &           && 1.999 859 & 0.318 414 & 0.407 324 \\
3.6  &1.738 111        &0.004 023  &           && 1.999 608 & 0.238 969 & 0.367 027 \\
3.0  &1.370 511  &           &           && 1.997 109 &           & 0.222 222 \\
     &1.119 3$^{a}$    &           &           &           &           &        \\
     &1.367 9$^{b}$    &           &           &           &           &           \\
     &1.370 511$^{c}$  &           &           &           &           &          \\
     &1.367 99$^{d}$  &           &           &           &           &           \\
     &1.370 5$^{e}$    &           &           &           &           &           \\
     &1.370 511$^{f}$\\
2.7  &1.051 349  &           &           && 1.992 388 &           & 0.092 647          \\
2.5  &0.751 657  &           &           && 1.985 668 &           &           \\
     &0.751 657$^{c}$  &           &           &           &           &        \\
     &0.751 6$^{e}$\\
2.4  &0.565 424  &           &           && 1.980 407 &           &           \\
2.35 &0.461 133  &           &           && 1.977 112 &           &           \\
2.308&0.367 050  &           &           && 1.973 932 &           &           \\
2.18 &0.038 236  &           &           && 1.961 339 &           &           \\
\hline\hline\\
\multicolumn{6}{l}{$^{a}$Ref.\cite{pat04}; $^{b}$Ref.\cite{aru06}; $^{c}$Ref.\cite{mont13}; $^{d}$Ref.\cite{yak11};  $^{e}$Ref.\cite{flo08}; $^{f}$Ref.\cite{mont15}}&\\
\end{tabular}
\end{center}
\end{table*}
\begin{table}[tbp]
\caption {\rm {Energy values of $ Li^{+} $ and $ Li^{2+} $ confined in a spherical cavity of radius R. The uncertainty of the calculated energy values is of the order of $10^{-6}$ a.u.}}
\begin{center}
\begin{tabular}{l l l l l l l l}\\
\hline\hline\vspace{-0.2cm}\\
 $R$&\multicolumn{3}{c}{$-E$ (in a.u.) of $Li^{+}$}&&\multicolumn{3}{c}{$-E~$ (in a.u.) of  $Li^{2+}$}\\
  \cline{2-4}\cline{6-8}\vspace{-0.2cm}\\
(a.u.)& $1s2s$ & $1s3s$ & $1s4s$ && $1s$ &$ 2s $&$ 2p $\\
\hline\vspace{-0.2cm}\\
25.0  & 5.110 716& 4.751 691& 4.636 444&& 4.500 000& 1.125 000& 1.125 000\\
20.0  &          &          & 4.635 162&& 4.500 000& 1.125 000& 1.125 000\\
10.0  &          & 4.739 985& 4.487 763&& 4.500 000& 1.125 000& 1.125 000\\
7.0   & 5.109 727&          &          && 4.500 000& 1.124 946& 1.124 978\\
6.0   &          &          & 3.735 358&& 4.500 000& 1.124 472& 1.124 772\\
5.0   & 5.090 063& 4.307 657& 3.106 372&& 4.500 000& 1.120 491& 1.122 942\\
      & 5.090 063$^{a}$ & 4.307 657$^{a}$ & 3.106 266$^{a}$ \\
      & 5.090 0$^{b}$\\
4.5   &          &          & 2.580 049&& 4.500 000& 1.112 665& 1.119 183\\
4.3   &          &          & 2.312 117&& 4.500 000& 1.106 787& 1.116 299\\
4.0   &          &          & 1.824 382&& 4.500 000& 1.092 743& 1.109 295\\
3.5   &          &          & 0.675 824&& 4.499 997& 1.043 812& 1.084 399\\
3.22  &          &          & 0.013 485&& 4.499 988& 0.990 455& 1.056 975\\
3.0   & 4.771 861& 2.420 112&          && 4.499 961& 0.925 512& 1.023 546\\
      & 4.771 861$^{a}$ & 2.420 112$^{a}$ \\
      & 4.771 9$^{b}$\\
      & 4.769 21$^{c}$\\      
2.8   &          & 1.915 481&          && 4.499 888& 0.839 733& 0.979 489\\
2.7   &          & 1.613 015&          && 4.499 811& 0.783 714& 0.950 813\\
2.6   &          & 1.269 561&          && 4.499 683& 0.716 432& 0.916 479\\
2.5   &          & 0.877 992&          && 4.499 471& 0.635 439& 0.875 301\\
2.4   &          & 0.429 563&          && 4.499 118& 0.537 681& 0.825 811\\
2.36  &          & 0.231 959&          && 4.498 919& 0.493 057& 0.803 291\\
2.32  &          & 0.022 725&          && 4.498 677& 0.444 864& 0.779 015\\
2.0   & 3.508 574&          &          && 4.493 495&          & 0.500 000\\
      & 3.508 5$^{b}$\\
      & 3.503 43$^{c}$\\
1.7   & 2.365 395&          &          && 4.472 446&          & 0.008 714\\
1.6   & 1.785 268&          &          && 4.455 916&          &          \\
      & 1.785 2$^{b}$\\
      & 1.772 60$^{c}$\\
1.5   & 1.052 329&          &          && 4.429 849&          &          \\
1.45  & 0.613 285&          &          && 4.411 669&          &          \\
1.446 & 0.575 736&          &          && 4.410 030&          &          \\
1.4452& 0.568 179&          &          && 4.409 699&          &          \\
1.39  & 0.007 806&          &          && 4.383 701&          &          \\
\hline\hline\\
\multicolumn{6}{l}{$^{a}$Ref.\cite{mont15}; $^{b}$Ref.\cite{flo08}; $^{c}$Ref.\cite{yak11}}\\
\end{tabular}
\end{center}
\end{table}
\begin{table}[tbp]
\caption {\rm {Energy values of $ Be^{2+} $ and $ Be^{3+} $ confined in a spherical cavity of radius R. The uncertainty of the calculated energy values is of the order of $10^{-6}$ a.u.}}
\begin{center}
\begin{tabular}{l l l l l l l l}\\
\hline\hline\vspace{-0.2cm}\\
$R$ & \multicolumn{3}{c}{$-E$ (in a.u.) of  $ Be^{2+} $}&&\multicolumn{3}{c}{$-E~$ (in a.u.) of $ Be^{3+} $}\\
\cline{2-4}\cline{6-8}\vspace{-0.2cm}\\
(a.u.)& $1s2s$ & $1s3s$ & $1s4s$ && $1s$ &$ 2s $&$ 2p $\\
\hline\vspace{-0.2cm}\\
25.0  & 9.297 159& 8.545 974& 8.299 073&& 8.000 000& 2.000 000& 2.000 000\\
10.0  &          & 8.545 587&          && 8.000 000& 2.000 000& 2.000 000\\
7.0   & 9.297 156&          &          && 8.000 000& 2.000 000& 2.000 000\\
5.0   &          &          & 7.248 942&& 8.000 000& 1.999 794& 2.000 000\\
4.0   &          &          & 6.174 433&& 8.000 000& 1.996 004& 1.999 914\\
3.5   &          &          & 5.165 647&& 8.000 000& 1.984 240& 1.998 213\\
3.0   &          &          & 3.513 852&& 8.000 000& 1.942 654& 1.992 649\\
2.7   &          &          & 1.974 703&& 7.999 997& 1.879 696& 1.972 081\\
2.5   &          &          & 0.591 075&& 7.999 988& 1.804 899& 1.940 121\\
2.4   &          &          & 0.178 918&& 7.999 976& 1.752 070& 1.901 753\\
2.39  &          &          & 0.146 069&& 7.999 974& 1.746 072& 1.874 569\\
2.36  &          &          & 0.042 465&& 7.999 968& 1.727 203& 1.871 481\\
2.0   & 8.410 046& 3.0161 60&          && 7.999 602& 1.355 819& 1.861 766\\
      & 8.410 0$^{a}$\\
      & 8.405 58$^{b}$\\
1.8   &          & 1.140 233&          && 7.998 432& 0.955 877& 1.671 201\\
1.71  &          & 0.040 090&          && 7.997 122& 0.698 286& 1.468 108\\
1.5   & 6.619 594&          &          && 7.988 436&          & 1.338 373\\
1.2   & 3.634 049&          &          && 7.921 629&          & 0.888 888\\
      & 3.634 0$^{a}$\\
      & 3.623 81$^{b}$\\
1.1   & 1.857 693&          &          && 7.854 559&          &          \\
1.08  & 1.426 571&          &          && 7.835 584&          &          \\
1.06  & 0.965 100&          &          && 7.814 189&          &          \\
1.055 & 0.844 689&          &          && 7.808 428&          &          \\
1.053 & 0.795 939&          &          && 7.806 075&          &          \\
1.0525& 0.783 698&          &          && 7.805 482&          &          \\
1.023 & 0.022 281&          &          && 7.767 174&          &          \\
\hline\hline\\
\multicolumn{6}{l}{$^{a}$Ref.\cite{flo08}; $^{b}$Ref.\cite{yak11}}\\
\end{tabular}
\end{center}
\end{table}
\begin{table}[tbp]
\caption {\rm {Energy values of $B^{3+}$ and $B^{4+}$ confined in a spherical cavity of radius R. The uncertainty of the calculated energy values is of the order of $10^{-6}$ a.u.}}
\begin{center}
\begin{tabular}{l l l l l l l l}\\
\hline\hline\vspace{-0.2cm}\\
$R$&\multicolumn{3}{c}{$-E$ (in a.u.) of $B^{3+}$}&&\multicolumn{3}{c}{$-E~$ (in a.u.) of $B^{4+}$}\\
\cline{2-4}\cline{6-8}\vspace{-0.2cm}\\
(a.u.)& $1s2s$ & $1s3s$ & $1s4s$&& $1s$ &$ 2s $&$ 2p $\\
\hline\vspace{-0.2cm}\\
15.0  & 14.733 877& 13.451 496& 13.024 215&& 12.500 000& 3.125 000& 3.125 000\\
10.0  &           &           & 13.016 808&& 12.500 000& 3.125 000& 3.125 000\\
5.0   & 14.733 832& 13.387 693&           && 12.500 000& 3.124 994& 3.125 000\\
      & 14.733 7$^{a}$\\
4.0   &           &           & 11.477 388&& 12.500 000& 3.124 678& 3.124 865\\
3.5   &           &           & 10.614 148&& 12.500 000& 3.122 881& 3.124 078\\
3.0   &           &           &  9.141 038&& 12.500 000& 3.112 475& 3.119 283\\
2.7   &           &           &  7.754 814&& 12.500 000& 3.090 738& 3.108 843\\
2.6   &           &           &  7.184 535&& 12.500 000& 3.077 567& 3.102 387\\
2.5   &           &           &  6.488 416&& 12.500 000& 3.059 655& 3.093 513\\
2.4   &           &           &  5.736 959&& 12.500 000& 3.035 397& 3.081 376\\
2.3   &           &           &  4.846 007&& 12.499 999& 3.002 662& 3.064 856\\
2.2   &           &           &  3.814 882&& 12.499 997& 2.958 618& 3.042 464\\
2.1   &           &           &  2.613 888&& 12.499 992& 2.899 479& 3.012 220\\
2.0   & 14.279 084&  9.150 376&  1.208 708&& 12.499 981& 2.820 155& 2.993 362\\
      & 14.279 0$^{a}$\\
1.9   &           &           &  0.350 119&& 12.499 955& 2.713 756& 2.916 733\\
1.89  &           &           &  0.284 693&& 12.499 951& 2.701 291& 2.889 888\\
1.86  &           &           &  0.080 654&& 12.499 936& 2.661 613& 2.910 316\\
1.8   &           &  7.496 836&           && 12.499 891& 2.570 867& 2.843 184\\
1.5   &           &  3.348 166&           && 12.498 529& 1.765 108& 2.431 393\\
1.352 &           &  0.012 387&           && 12.494 858& 0.988 168& 2.039 561\\
1.2   & 10.703 658&           &           && 12.481 932&          & 1.388 889\\
      & 10.703 6$^{a}$\\
1.0   & 7.186 882&           &           && 12.410 425&          &          \\
      & 7.186 8$^{a}$\\
0.95  &  5.815 658&           &           && 12.367 639&          &          \\
0.9   &  4.142 045&           &           && 12.305 136&          &          \\
0.87  &  2.959 152&           &           && 12.254 638&          &          \\
0.85  &  2.082 734&           &           && 12.214 072&          &          \\
0.84  &  1.615 349&           &           && 12.191 393&          &          \\
0.83  &  1.127 117&           &           && 12.166 951&          &          \\
0.828 &  1.026 869&           &           && 12.161 839&          &          \\
0.8274&  0.996 623&           &           && 12.160 291&          &          \\
0.81  &  0.083 461&           &           && 12.112 224&          &          \\
\hline\hline\\
\multicolumn{6}{l}{$^{a}$Ref.\cite{flo08}}\\
\end{tabular}
\end{center}
\end{table}

\begin{table}[tbp]
\caption {\rm {Confinement radii ($R_{th}$) in a.u. and corresponding pressures ($P_{th}$) in Pa when the bound $1sns~(^{1}S^e)~[n=1-3]$ and $1sn's~(^{3}S^e)~[n'=2-3]$ states of different ions ($Z = 2-5$) cross the respective first one-electron ionization threshold. Conversion factor : 1 a.u. of pressure = 2.9421912(13) Pa. The notation $x(y)$ indicates $x \times 10^{y}$.}}
\begin{center}
\begin{tabular}{l l l l l l l l l}\\
\hline\vspace{-0.3cm}\\
 & \multicolumn{2}{c}{$He$}&\multicolumn{2}{c}{$Li^+$}&\multicolumn{2}{c}{$Be^{2+}$}&\multicolumn{2}{c}{$B^{3+}$}\\
  \cline{2-9}\vspace{-0.2cm}\\
State & \multicolumn{1}{c}{$R_{th}$}&\multicolumn{1}{c}{$P_{th}$}& \multicolumn{1}{c}{$R_{th}$}&\multicolumn{1}{c}{$P_{th}$}& \multicolumn{1}{c}{$R_{th}$}&\multicolumn{1}{c}{$P_{th}$}& \multicolumn{1}{c}{$R_{th}$}&\multicolumn{1}{c}{$P_{th}$}\\
\hline\vspace{-0.2cm}\\
$1s^2(^1S^e)$ & 1.537 &0.1727(13)& 0.909 & 0.1485(14) & 0.646 & 0.5940(14) & 0.5013 & 0.1655(15)\\
$1s2s(^1S^e)$ & 5.33 & 0.7493(08)  & 2.87 & 0.2191(10) &1.955 & 0.1499(11) & 1.485 & 0.5446(11)\\
$1s3s(^1S^e)$ & 11.8 & 0.1984(04)& 6.195 & 0.5669(04) &4.19  & 0.2794(05) & 3.165 & 0.3699(05) \\
\hline\vspace{-0.2cm}\\
$1s2s(^3S^e)$ & 4.6  & 0.4796(09) & 2.6 & 0.7090(10) & 1.82 & 0.3431(11) & 1.403 & 0.1042(12)\\
$1s3s(^3S^e)$ & 10.5 & 0.3345(04) & 5.75 & 0.1018(05) & 3.98 & 0.2941(05) &3.043 &  0.3847(06)\\
\hline\hline
\end{tabular}
\end{center}
\end{table}

\begin{table}[tbp]
\caption {\rm {Critical radii ($R_c$) in a.u. and corresponding pressures ($P_c$) in Pa for destabilization of the bound $1sns~(^{1}S^e)~[n=1-4]$ and $1sn's~(^{3}S^e)~[n'=2-4]$ states of different ions ($Z = 2-5$). Conversion factor : 1 a.u. of pressure = 2.9421912(13) Pa. The notation $x(y)$ indicates $x \times 10^{y}$.}}
\begin{center}
\begin{tabular}{l l l l l l l l l}\\
\hline\vspace{-0.3cm}\\
 & \multicolumn{2}{c}{$He$}&\multicolumn{2}{c}{$Li^+$}&\multicolumn{2}{c}{$Be^{2+}$}&\multicolumn{2}{c}{$B^{3+}$}\\
  \cline{2-9}\vspace{-0.2cm}\\
State & \multicolumn{1}{c}{$R_c$}&\multicolumn{1}{c}{$P_c$}& \multicolumn{1}{c}{$R_c$}&\multicolumn{1}{c}{$P_c$}& \multicolumn{1}{c}{$R_c$}&\multicolumn{1}{c}{$P_c$}& \multicolumn{1}{c}{$R_c$}&\multicolumn{1}{c}{$P_c$}\\
\hline\vspace{-0.2cm}\\
$1s^2(^1S^e)$ $^a$ & 1.11 & 0.1544(14)  & 0.69 & 0.1658(15) & 0.502  & 0.8097(15) & 0.394  & 0.2713(16)\\
$1s2s(^1S^e)$ & 2.308 & 0.1702(12)& 1.4452 & 0.1051(13) & 1.0525 & 0.3645(13) & 0.8274 & 0.9412(13)\\
$1s3s(^1S^e)$ & 3.208 & 0.1607(11) & 1.95 & 0.1160(12) & 1.435 & 0.3537(12) &  1.121 & 0.9439(12)\\
$1s4s(^1S^e)$ & 3.621 & 0.5654(10) & 2.263 & 0.3030(11) & 1.653 & 0.9480(11) & 1.302 & 0.2304(12) \\
\hline\vspace{-0.2cm}\\
$1s2s(^3S^e)$ &2.18 &0.2431(12)&1.39 & 0.1350(13) & 1.023  &0.4388(13) & 0.81 &0.1082(14)\\
$1s3s(^3S^e)$ &3.6 &0.5961(10) &2.32 &0.2372(11) & 1.71 & 0.6709(11) & 1.352 & 0.1556(12)\\
$1s4s(^3S^e)$ &5.1 &0.1349(09) &3.22 &0.4690(09) &2.36 &0.1208(10) & 1.86 &0.2669(10) \\
\hline\hline
$^a$Ref.\cite{bhat13}
\end{tabular}
\end{center}
\end{table}

\begin{figure}[!th]
\centerline{\epsfxsize=10.0cm\epsfysize=12.0cm\epsfbox{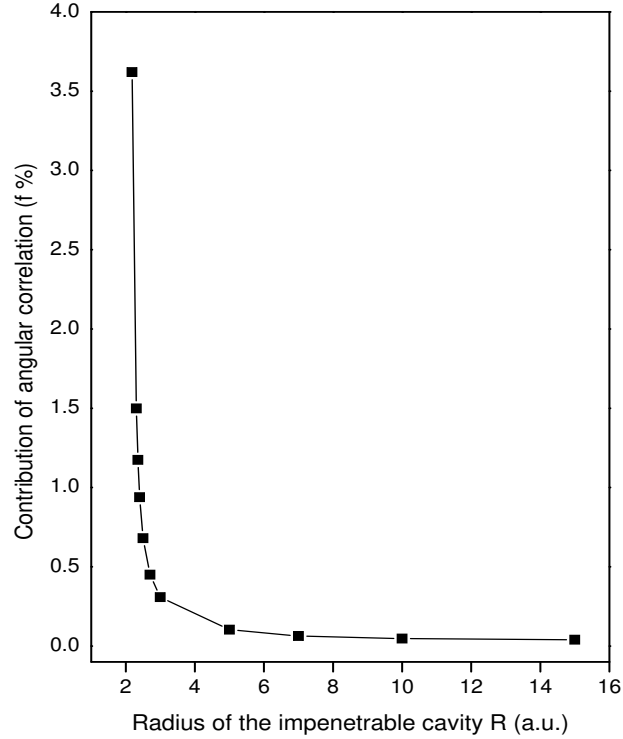}}
\caption{Variation of the percentage contribution of the angular correlation to the energy of $ 1s2s $ ($ ^3S^e $) state of helium \textit{w.r.t.} the confinement radius $R$ (a.u.).}
\end{figure}

\begin{figure}[!th]
\centerline{\epsfxsize=10.0cm\epsfysize=12.0cm\epsfbox{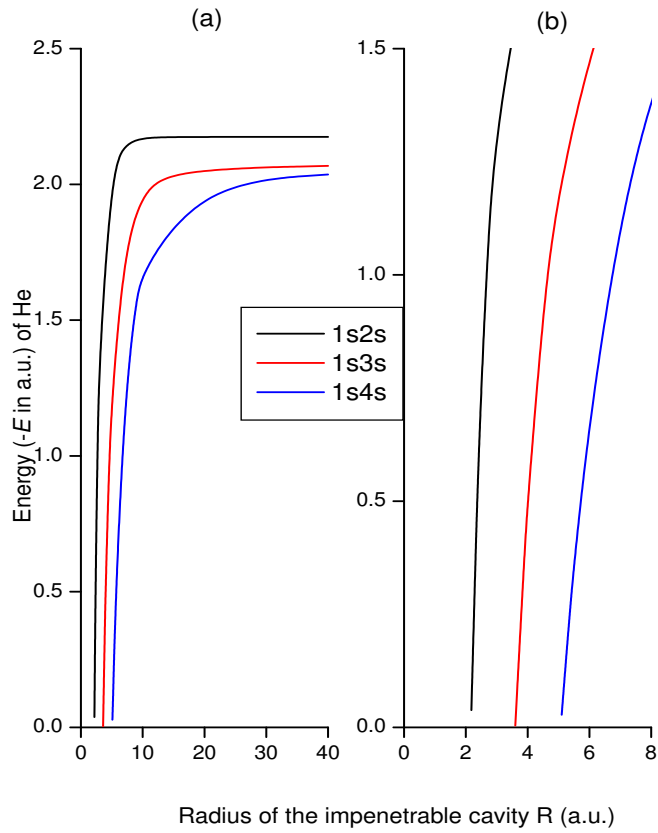}}
\caption{(a) Variation of eigen energies ($-E$) of bound $1sns$($^3S^e$) [$n=2-4$] states of helium \textit{w.r.t.} the confinement radius $R$ (a.u.). (b) Enlarged view for $1sns$($^3S^e$) states near the destabilization region.}
\end{figure}

\begin{figure}[!th]
\centerline{\epsfxsize=10.0cm\epsfysize=12.0cm\epsfbox{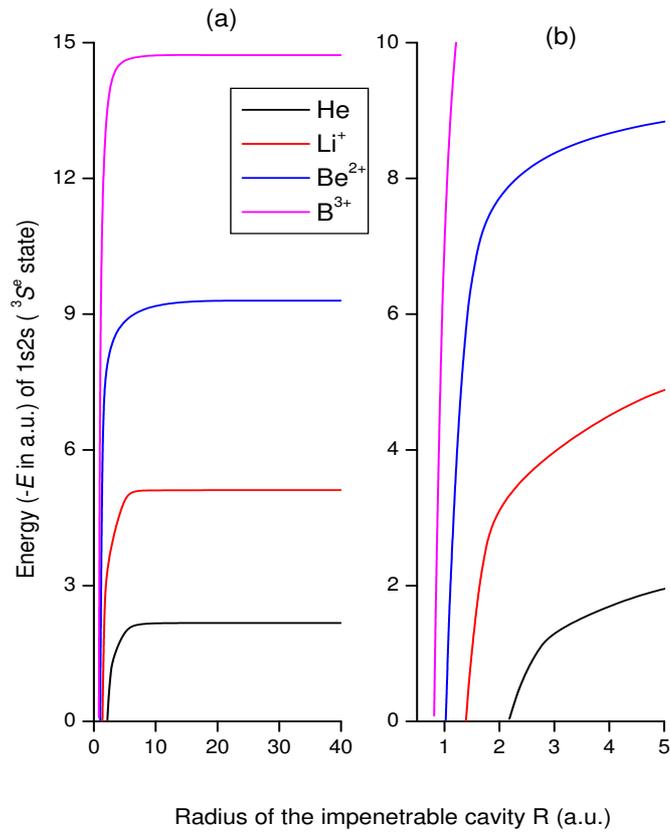}}
\caption{(a) Variation of eigen energies ($-E$) of $1s2s$($^3S^e$) states of of $He, Li^+, Be^{2+}$ and $B^{3+}$ with radius $ R $ of the impenetrable cavity. (b) Enlarged view near the destabilization region.}
\end{figure}

\begin{figure}[!th]
\centerline{\epsfxsize=12.0cm\epsfysize=10.0cm\epsfbox{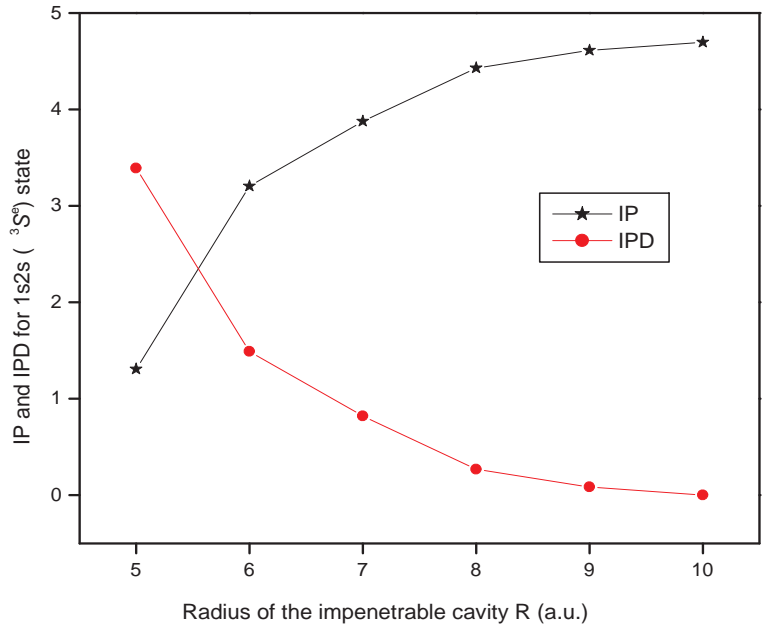}}
\caption{Variation of ionization potential (IP) and ionization potential depression (IPD) for $1s2s$($^3S^e$) state of helium \textit{w.r.t.} the confinement radius $R$ (a.u.).}
\end{figure}

\begin{figure}[!th]
\centerline{\epsfxsize=12.0cm\epsfysize=10.0cm\epsfbox{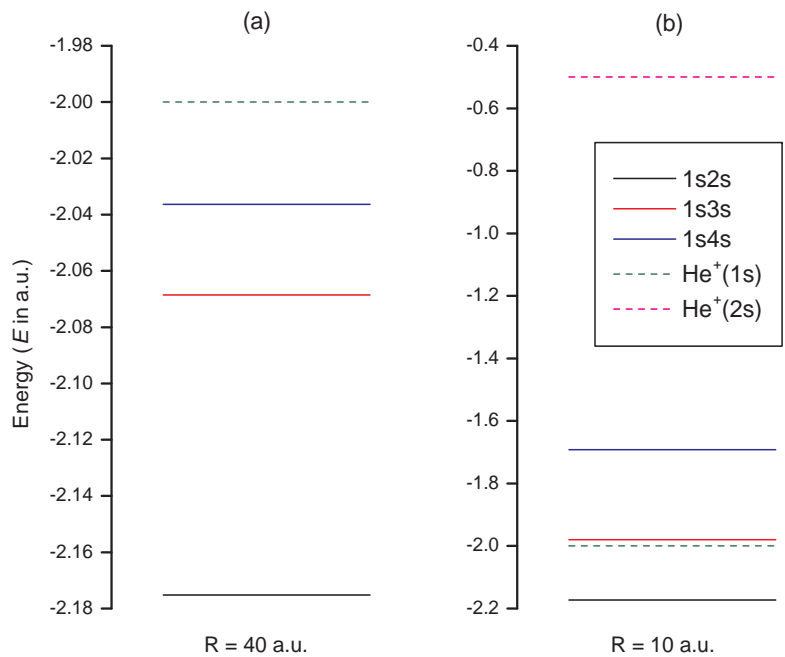}}
\caption{(a) Energy level diagram of helium at $ R=40.0 $ a.u. where the $1sns$($^3S^e$) [$n=2-4$] states lie below $ N=1 $ ionization threshold of $ He^+ $. (b) Energy level diagram of helium at $ R=10.0 $ a.u. where $1s3s$ and $1s4s$ ($^3S^e$) states have moved above $ N=1 $ ionization threshold of $ He^+ $.}
\end{figure}

\begin{figure}[!th]
\centerline{\epsfxsize=10.0cm\epsfysize=12.0cm\epsfbox{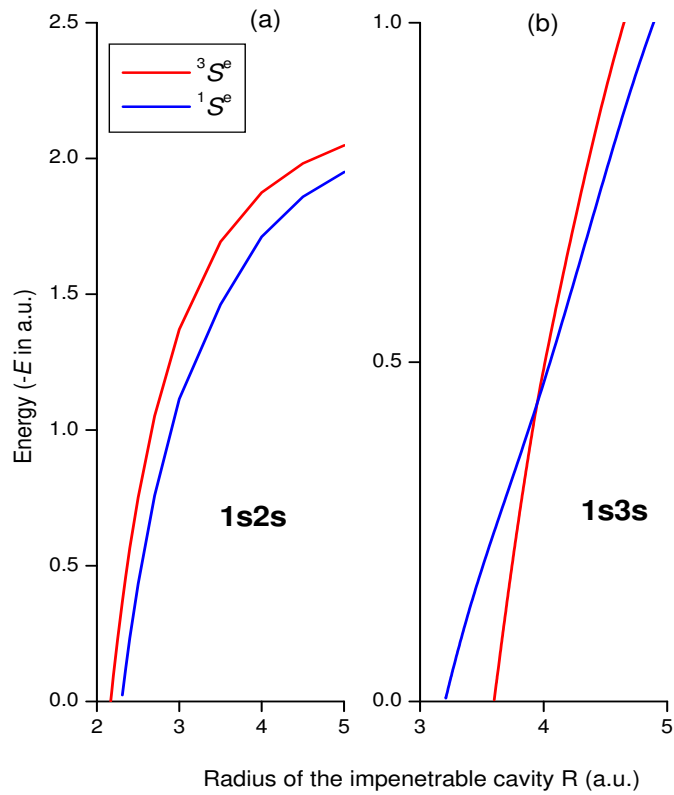}}
\caption{(a) Variation of $1s2s$($^3S^e$) and $1s2s$($^1S^e$) states of helium \textit{w.r.t.} the confinement radius $R$ (a.u.). (b) Variation of $1s3s$($^3S^e$) and $1s3s$($^1S^e$) states of helium \textit{w.r.t.} the confinement radius $R$ (a.u.) where \textit{level-crossing} occurs.}
\end{figure}

\begin{figure}[!th]
\centerline{\epsfxsize=10.0cm\epsfysize=12.0cm\epsfbox{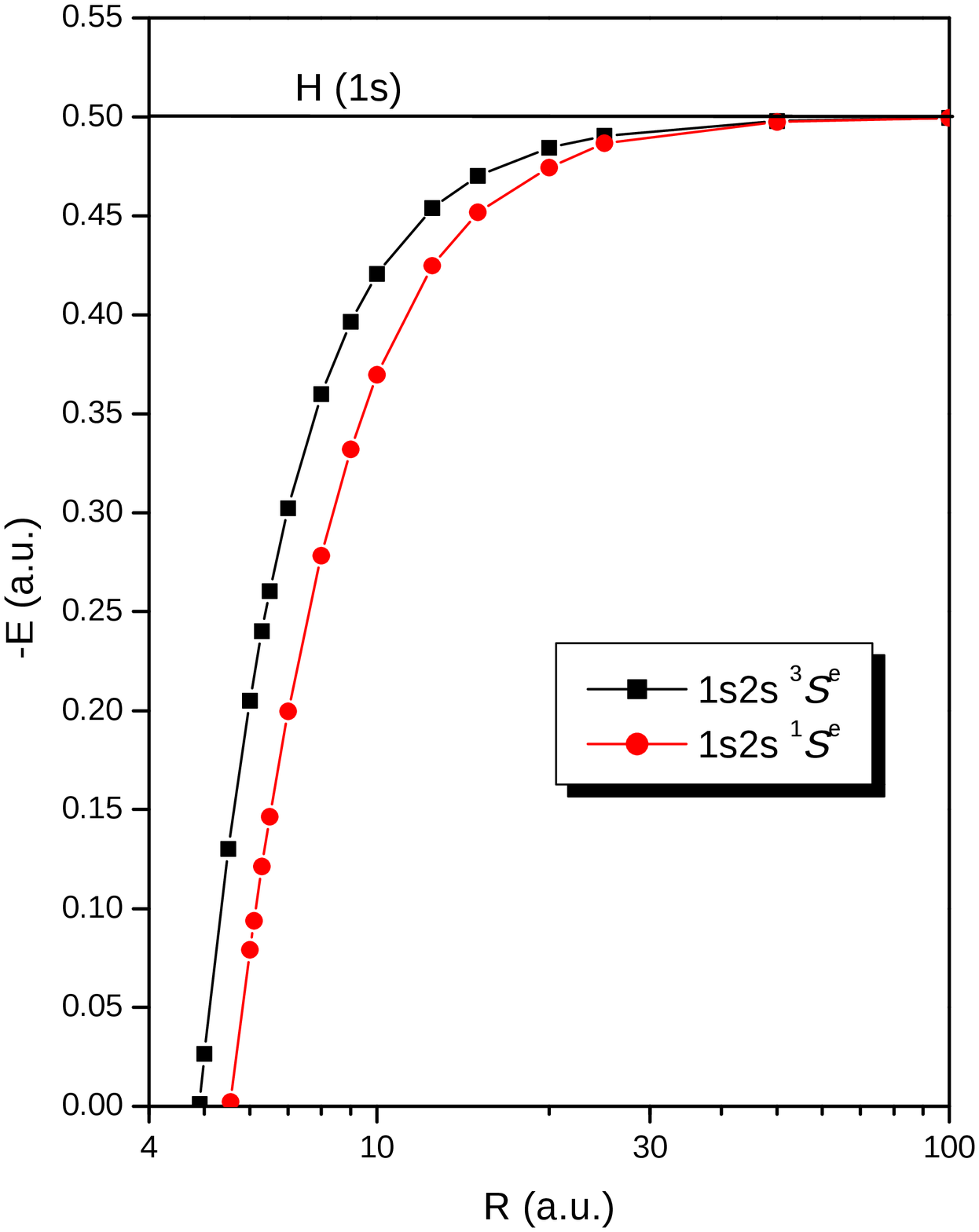}}
\caption{Variation of eigen energies ($-E$) of bound $1s2s$($^{1,3}S^e$) states of $H^-$ \textit{w.r.t.} the confinement radius $R$ (a.u.).}
\end{figure}


\begin{thebibliography}{80}
\bibitem{can08} S. Canuto (\textit{ed}.) Solvation Effects on Molecules and Biomolecules, Computational Methods and Applications (Berlin: Springer) (2008).
\bibitem{sab09} J. Sabin and E. Brandas (\textit{Ed.}) and prefaced by S. A. Cruz, Adv. Quantum Chem. \textbf{58} and \textbf{59}(2009).
\bibitem{sen14} K. D. Sen (\textit{Ed.}) Electronic Structure of Quantum Confined Atoms and Molecules, (Switzerland: Springer) (2014)
\bibitem{sil09} A.N. Sil, S. Canuto, and P. K. Mukherjee, Adv. Quantum Chem. \textbf{58}, 115 (2009) and references therein.
\bibitem{bhat15} S. Bhattacharyya, J. K. Saha and T. K. Mukherjee, Phys. Rev. A \textbf{91}, 042515 (2015) and references therein.
\bibitem{xu96} Y. B. Xu, M. Q. Tan, and U. Becker, Phys. Rev. Lett. \textbf{76}, 3538 (1996).
\bibitem{gen10} M. Genkin and E. Lindroth, Phys. Rev. B 81 (2010) 125315.
\bibitem{nor08} D. J. Norris, A. L. Efros and S. C. Erwin Science 319 (2008) 1776.
\bibitem{jac97} P. A. Jacobs, Carboniogenic Activity of Zeolites (Amsterdam: Elsevier) (1997).
\bibitem{wal00} C.A. Walsh, J. Yuan, and L.M. Brown, Philos. Mag. B \textbf{80}, 1507 (2000).
\bibitem{gul99} T. Guillot,  Planet Space Sci. \textbf{47}, 1183 (1999).
\bibitem{mic37} A. Michels, de J Boer and A. Bijl, Physica \textbf{4} 981 (1937).
\bibitem{som38} A. Sommerfeld and H. Welker, Ann. Phys. \textbf{32}, 56 (1938).
\bibitem{mont0211} H. E. Montgomery, Chem. Phys. Lett. 352 529 (2002); Eur. J. Phys. 32 1275 (2011).
\bibitem{lau02} C. Laughlin, B. L. Burrows and M. Cohen, J. Phys. B \textbf{35}, 710 (2002).
\bibitem{lau04} C. Laughlin, J. Phys. B \textbf{37}, 4085 (2004).
\bibitem{cif09} H. Ciftci, R. L. Hall and N. Saad, Int. J. Quan. Chem. \textbf{109}, 931 (2009).
\bibitem{ste08} L. Stevanovic and K. D. Sen, Publ. Astron. Obs. \textbf{84}, 111 (2008).
\bibitem{aqu9511} N. Aquino, Int. J. Quan. Chem. \textbf{54}, 107 (1995); Rev. Mex. de
Fis. \textbf{57}, 65 (2011).
\bibitem{bhat13} S. Bhattacharyya, J. K. Saha, P. K. Mukherjee and T. K. Mukherjee, Phys. Scr. \textbf{87}, 065305 (2013) and references therein.
\bibitem{gim67} B. M. Gimarc, J. Chem. Phys. \textbf{47}, 5110 (1967). 
\bibitem{lud79} E. V. Ludena and M. Gregori, J. Chem. Phys. \textbf{71}, 2235 (1979).
\bibitem{mar91} J. L. Marin and S. A. Cruz, J. Phys. B \textbf{24}, 2899 (1991).
\bibitem{jos92} C. Joslin and S. Goldman, J. Phys. B \textbf{25}, 1965 (1992).
\bibitem{aqu03} N. Aquino, A. Flores-Riveros, and J. F. Rivas-Silva, Phys. Lett. A \textbf{307}, 326 (2003).
\bibitem{aru06} A. Banerjee, C. Kamal, A. Chowdhury, Phys. Lett. A \textbf{350}, 121 (2006).
\bibitem{flo08} A. Flores-Riveros, and A. Rodriguez-Contreras, Phys. Lett. A \textbf{372}, 6175 (2008).
\bibitem{lau09} C. Laughlin, and S. I. Chu, J. Phys. A: Math. Theor. \textbf{42}, 265004 (2009).
\bibitem{flo10} A. Flores-Riveros, N. Aquino, and  H. E. Montgomery Jr., Phys. Lett. A \textbf{374}, 1246 (2010).
\bibitem{mont13} H. E. Montgomery Jr., and V. I. Pupyshev, Phys. Lett. A \textbf{377}, 2880 (2013).
\bibitem{pat04} S. H. Patil and Y. P. Varshni, Can. J. Phys. \textbf{82} 647 (2004).
\bibitem{yak11} Y. Yakar, B. Cakir, A. Ozmen, Int. J. Quant. Chem. \textbf{111}, 4139 (2011). 
\bibitem{mont15} H. E. Montgomery Jr., and V. I. Pupyshev, Theo. Chem. Acc. \textbf{134}, 1598 (2015).
\bibitem{sen05} K. D. Sen, J. Chem. Phys. \textbf{122}, 194324 (2005).
\bibitem{dut15} S. Dutta, J. K. Saha, Phys. Plas. \textbf{22}, 062103 (2015);(2015).
\bibitem{tkm94} T. K. Mukherjee, and P. K. Mukherjee, Phys. Rev. A \textbf{50}, 850 (1994).
\bibitem{nel65} J. A. Nelder, and R. Mead, Comput. J. \textbf{7}, 308 (1965).
\bibitem{car09} E. A. Carrillo-Delgado, I. Rodriguez-Vargas, and S. J. Vlaev, PIERS Online \textbf{5}, 137 (2009) and references therein.
\bibitem{cap92} F. Capasso, C. Sirtori, J. Faist, D. L. Sivco, S. G. Chu, and A. Y. Cho, Nature (London) \textbf{358}, 565 (1992).
\end{thebibliography}
\end{document}